# Evaluation of Image Registration for measuring deformation fields in soft tissue mechanics

Running head title: Deformation fields with Image Registration

Ondřej Lisický[1*], Stéphane Avril[2], Bastien Eydan[2], Baptiste Pierrat[2], Jiří Burša[1]

[1] *Institute of Solid Mechanics, Mechatronics and Biomechanics, Brno University of Technology, Czech Republic*

[2] *Mines Saint-Etienne, University of Lyon, University Jean Monnet, INSERM, Saint-Etienne, France*

*Corresponding author*

*E-mail: 161238@vutbr.cz*

Phone: +420 605 292 174

# Abstract

High-fidelity biomechanical models usually involve the mechanical characterization of biological tissues using experimental methods based on optical measurements. In most experiments, strains are evaluated based on displacements of a few markers and represents an average within the region of interest (ROI). Full-field measurements may improve description of non-homogeneous materials such as soft tissues. The approach based on non-rigid Image Registration is proposed and compared with standard Digital Image Correlation (DIC) on a set of samples, including (i) complex heterogeneous deformations with sub-pixel displacement, (ii) a typical uniaxial tension test of aorta, and (iii) an indentation test on skin. The possibility to extend the ROI to the whole sample and the exploitation of a natural tissue pattern represents the main assets of the proposed method whereas the results show similar accuracy as standard DIC when analysing sub-pixel deformations. Therefore, displacement and strain fields measurement based on Image Registration is very promising to characterize heterogeneous specimens with irregular shapes and/or small dimensions, which are typical features of soft biological tissues.

# Keywords

# 1 Introduction

The expanding use of computational modelling in biomechanics [1–3], e.g., for the design of medical devices and possibly for clinical prognosis in the near future, requires appropriate knowledge of tissue behaviour. Accordingly, patient-specific medical imaging like computed tomography, magnetic resonance imaging (MRI) or intravascular ultrasound are needed for assessing the mechanical behaviour of biological tissues. MRI is a very versatile method enabling strain investigation of various tissues under load and had already been applied for tissues like myocardium [4,5], carotid artery [6], spine intervertebral disc [7] or brain [8]. Tissue deformation in the latter examples was examined using the Image Registration (IR) technique, which is normally used to align medical images from multiple modalities although its versatility enables various other applications. However, the information about *in vivo* strain cannot be easily used for constitutive model characterization, especially for soft biological tissues, often exhibiting a nonlinear behaviour under large deformations.

Various approaches used to characterise mechanical properties of soft tissues were recently reviewed [9]. The choice of a suitable method depends mostly on the available laboratory equipment and the specific application; for instance, biaxial testing requires specific testing machines and large sample specimens which may not be available with some tissues [10,11]. Another example is the inflation-extension test requiring cylinder-like specimens [12], thus hardly applicable for the aortic valve or atherosclerotic plaque [13]. Most of the experimental studies investigating mechanical properties of biological tissues assume a uniform response in the analysed area and global characteristics are obtained by tracking the displacement of e.g., few markers resulting in average deformation of the specimen [9,14] With this approach, it is straightforward to derive stress-strain curves which can subsequently be processed to identify a constitutive model.

However, the assumption of homogeneous strain and stress fields in the analysed area may not be met in soft tissue mechanics. The error is then transferred into the computational model and can contribute to high variation among studies with heterogeneous samples [15] and a global trend for its minimization can be seen [14]. Regional characterization of the sample deformation during the experiment may help to capture heterogeneous strain fields and can be addressed by full-field methods like digital image correlation (DIC), moiré or speckle interferometry. DIC has been widely used in many biomechanical applications for decades, including e.g., human skin [16], thigh muscle [17], human ascending thoracic aneurysm wall [18], or recently for biaxial stress relaxation of tissue of vaginal mucosa [19]. An extension of this method enables analysis of surface deformation with 3D DIC [20,21] or even internal deformation of the volume with digital volume correlation (DVC) [22–24]. DIC has limitations influencing its outcomes [25] such as creation of the pattern [26] or noisy deformation field as a result of the local DIC. The latter issue is addressed by introducing global DIC enabling an investigation of continuous deformation field in combination with finite element method [27] or by introducing B-spline shape function [28,29]. However, a comparison between global FE-based DIC and local DIC [30] revealed that the local DIC outperforms its opponent when a subset size is no less than 11 pixels. Other approaches were proposed to obtain a deformation/strain field of a loaded sample describing its regional mechanical properties. Even though IR is mostly used in computer vision, Wang *et al.* [25] analysed similarities, differences and complementarity between DIC and IR. According to that study, both methods may benefit from their combination and synergy. Non-rigid B-spline IR is similar to the global DIC with a continuous deformation field, however, has less strict requirements for the pattern and even a natural surface might be sufficient. IR approach was firstly used to evaluate the compressibility of lung parenchyma in uniaxial tension test [31]. However, neither verification nor reproducibility analyses were described in that study, although the same methodology was applied several times [32,33]. Zhu et al. [34] recently compared both methodologies with promising results on synthetic data. Another approach was used by Miga *et al.* [35] who, similarly to the IR based methods, deformed one image to fit the shape of another using finite element simulations.

In summary, even though deformation fields are crucial for soft tissue biomechanics, the use of local DIC still presents some limitations. In this study, we propose to overcome these limitations by evaluating the benefits of IR for soft tissues biomechanics.

# 2 Material and methods

## 2.1 Full-field analysis

The composition of most soft tissues, with some exceptions like liver [33], does not show any significant surface features, and hence intensity-based IR was applied. The IR is highly dependent on the specific tissue and a customized solution might be required. Therefore, open-source Elastix [36,37] software was used to simplify the process, covering a large field of problems and ways of solution.

Sets of N recorded images (including the first undeformed image) were evaluated using the incremental approach based always on comparison of two consecutive images from the set. In the IR, the so-called moving image $I_M$ (from the actual, i.e. $i^{th}$ step of loading) is distorted via $\mathbf{T(x)}$ transformation into the reference (fixed) image $I_F$ (recorded in the next step i+1) to find the displacements $\mathbf{u(x)}$ spatially aligning $I_M$ with $I_F$ (see Fig. 1). The optimization problem minimizing a cost function was solved using Adaptive Stochastic Gradient Descent method which requires to set less parameters than a regular gradient descent algorithm as it adaptively predicts step size during optimization, and performs well for IR problems [38]. This method works especially well with a random image sampler, which selects randomly a user-specified number of pixels from the images and offers high effectiveness and speed. The quality of the alignment during optimization is assessed with a similarity measure (also called cost function or error function) which is rather intensity-based than geometrically based in an IR problem. The alignment of images from the same modality, being the objective of our procedure, works well with metrics like Sum of Squared Differences or Normalized Correlation Coefficient. Nevertheless, the Mutual Information was found most suitable here for all the samples. This metric assumes only a relationship between the probability distributions of the intensities of the fixed and the moving images and is more general than the two metrics mentioned above. Due to its generalization, this metric is suitable for both mono- and multi-modal image pairs.

Fig. 1

Due to a common nonlinear behaviour of soft tissue, a more flexible model is required. Hence, the B-spline transform model, allowing local deformation, was used instead of the Demons algorithm applied in [34]. Therefore, images $I_M$ and $I_F$ were overlayed with a deformable regularly spaced grid with B-spline control points, the distance between which can be user-specified. The number of control points influences the final number of parameters which are optimized and should be tuned to obtain reasonable results. Displacement of each pixel defined in the undeformed state of the specimen grid was then calculated using *transformix* [36,37]. Moreover, the continuous and smooth description of deformation by the B-splines results directly in a smooth displacement field making the evaluation of strains easier. Here, directional strains were obtained by calculating central differences of the displacement field using the *gradient* function of a NumPy package. Afterwards, components $E_{ij}$ of Green-Lagrange strain tensor were calculated as follows:

$$E_{xx} = \frac{1}{2}\left(2\frac{\partial u}{\partial x} + \left(\frac{\partial u}{\partial x}\right)^2 + \left(\frac{\partial v}{\partial x}\right)^2\right)$$

$$E_{yy} = \frac{1}{2}\left(2\frac{\partial v}{\partial y} + \left(\frac{\partial u}{\partial y}\right)^2 + \left(\frac{\partial v}{\partial y}\right)^2\right)$$

$$E_{xy} = \frac{1}{2}(\frac{\partial u}{\partial y} + \frac{\partial v}{\partial x} + \frac{\partial u}{\partial x}\frac{\partial u}{\partial y} + \frac{\partial v}{\partial x}\frac{\partial v}{\partial y})$$

where $u$ and $v$ represent displacements in $x$ and $y$ directions, respectively.

In IR problems, the data complexity can be influenced with the use of a multi-resolution strategy. Down-sampling of the image resolution in combination with smoothing (also denoted as pyramids) is often applied to the images, meaning that more iterations are required, increasing thus the time needed to obtain acceptable results.. The use of multiple pyramid levels is shown e.g., in [34] where the authors obtained reasonable results for at least 8 pyramids. Nevertheless, the multi-resolution strategy was not necessary in the present study and only a single magnitude of resolution was used without pixel size degradation. However, this strategy can be easily implemented with Elastix when, for instance, deformations between the acquired images are too large.

### 2.1.1 DIC

DIC is a well-established method and was used here for a comparison of the obtained full-field measurements from experiments using the Ncorr software (local DIC) [39]. In contrast to the proposed IR approach, here the resulting displacement field, also dependent on a subset size, needs to be smoothed to evaluate the strain field. Therefore, Ncorr uses a strain window with variable radius $r_\varepsilon$ to calculate displacement gradients. As a result, a user needs to perform multiple analyses to find an acceptable compromise in the resulting fields. Therefore, the comparison was done with a DIC setup evaluated only personally, although by following all suggestions from the user guide.

## 2.2 Synthetic data – heterogeneous deformations

Synthetic data are often used to investigate the performance of some approaches. For the full-field measurements, a set of samples was proposed in [40], intending to challenge DIC algorithms to see their accuracy in various scenarios. While most of the samples presented there simulate only rigid motion with varying levels of noise or distortion, there are cases (like sample 14 in [40]) introducing sub-pixel heterogeneous deformation and thus a strain field representing a real challenge for the algorithms, as stated by the authors. This sample is very often used in testing new modifications of DIC [41,42] and hence it was chosen also here to challenge the IR approach. As it is suited for DIC, it has a typical speckle pattern though with high noise, introducing uncertainties into the analysed displacements. Moreover, this sample has three variations L1, L3 and L5 differing in the applied displacement. The speckle pattern of the synthetic data together with other samples can be seen in Fig. 2.

Fig. 2

## 2.3 Porcine aorta – uniaxial tension

Although synthetic data provides a good basis to establish the accuracy of the methodology, it is of interest to see the performance with data typical in the field of biomechanics. Here, a uniaxial tension of porcine proximal aorta sample was tested till relatively high deformations up to 80 %. It is worth mentioning that the sample had been frozen after harvesting and was tested in air. Although this can affect the response of the tissue, the focus of this study is the evaluation of the performance of the optical method. A speckle pattern was carefully applied onto the intima surface of a dog-bone shaped circumferential specimen (width = 4 mm) using a black spray. An Instron 3343 tensile machine was used to stretch the sample with a jaw velocity of 1.3 mm/s. Images were acquired during the test with a Photron UX100 camera (resolution of 1280x1024) at a framerate of 5 i/s for further full-field analysis.

## 2.4 Skin sample – natural pattern

Skin mechanical characterization is an important topic of interest in biomechanics. Mechanical properties of each layer can be identified using an indentation test in combination with uni- or biaxial skin stretching. A possible way how to obtain their constitutive parameters is by using inverse analysis with full-field measurements. However, the low thickness makes it almost impossible to prepare a suitable speckle pattern for DIC. The skin sample used in this study was stained with Hematoxylin and eosin to highlight a natural pattern of the layers. As no artificial pattern was prepared, this sample was very challenging for both DIC and IR approaches. The samples are 55mm long, die-cut from abdominal skin of a 90-100 days old female pig. No pre-stretch was applied before indentation which was performed with stainless steel probe with 4 mm diametr up to 1.93 N. The camera resolution 6000x4000.

## 2.5 Performance validation

Results obtained with the synthetic data were compared with the prescribed displacement and strain functions. On the other hand, the performance of IR approach was compared with the DIC results for two other samples (aorta and skin) using different patterns. DIC is considered here as a ground-truth for comparison as it is a well-established method for full-field measurements. A pixel-to-pixel comparison of full-field measurements was assessed as a mean absolute percentage error (MAPE) at each loading step for displacements and strains using:

$$MAPE(y, \hat{y}) = \frac{1}{n} \sum_{i=1}^{n} \frac{|y_i - \hat{y}_i|}{|y_i|}$$

where $n$ is number of pixels within the analysed ROI, $\hat{y}_i$ is the evaluated value from IR of i-th pixel and $y_i$ is the corresponding reference (taken as true) value from DIC. Additionally, the performance of IR can be also investigated to verify the alignment. However, this is not a straightforward problem and, for instance, the similarity measure used for optimization should not be used. Therefore, the performance of IR was investigated via average structural similarity index measure (SSIM) introduced in [43]. The SSIM is a perceptual metric that uses an idea that pixels have strong interdependencies when they are spatially close and thus an important structural information of the object is compared. This measure falls between 0 and 1 where values close to 1 indicate the perfect structural match of both images. The SSIM is calculated locally comparing the fixed and moving images after deformation, with a 21x21 sliding Gaussian window and a standard deviation of 1 pixel..

# 3   Results

## 3.1  Synthetic data – small sub-pixel deformations

As already mentioned, this sample is challenging due to a sub-pixel heterogeneous deformation. The proposed IR approach was capable to reproduce the simulated deformation successfully for all variants. Moreover, it captured well even the highest deformation gradients. This type of data provides a robust way to assess quality of the approach. Analogous to DIC, the setup might change the resulting displacement field and thus the strain field. In the case of DIC, the output field is influenced by the subset size $r$ and strain window radius $r_\varepsilon$ [39]. The trade-off between setups is well illustrated for Ncorr in Table 1 where a higher smoothing (ROI radius $r$=30 and $r_\varepsilon$ =20) is needed to obtain an acceptable SD which, however, results in a higher average response. The same applies for IR as a considerable bias is presented with decreasing local error see Fig. **3**.

Fig. 3

Consequently, a compromise between the mean response accuracy and the introduced variation is needed. Therefore, three different B-spline control points spacings (15, 30 and 60 pixels) were tested to indicate the necessary user interface to obtain the most suitable field. Table 1 summarizes the error in the resulting horizontal displacements and strains for both the approaches – the proposed IR and the DIC algorithms published in [39,40]. While the standard deviations in Table 1 suggest a better performance with a higher spacing of IR, it is unfortunately accompanied by a higher bias from the prescribed deformation.

Table 1

Nevertheless, even the middle spacing showed results comparable to the DIC algorithms and was therefore used for evaluation with different setups L3 and L5 (according to [40]) while L1 setup showed the best response with the spacing of 60 showing a low bias in the mean response. Fig. 4 shows the obtained results for all the three configurations of synthetic data analysed by the IR approach. Here, the mean displacement response (average of vertical points at each horizontal level) was well captured and the same applies for the calculated strains.

Fig. 4

## 3.2 Uniaxial tension – large deformations with artificial speckle pattern

The porcine specimen was tested till rupture. Every 10[th] image only was used for the analysis by both DIC and IR. The setup of DIC was adjusted multiple times until a trade-off between distribution of the displacement and strain fields was found. The subset radius had only a small impact on the resulting displacement distribution with maximum MAPE ~ 1 % (r = 10 vs r = 30). Similar results were obtained also with different strain window radius for smoothing, with maximum MAPE ~ 3 % between DIC strain fields with $r_\varepsilon = 5$ and 10. The final analysis was performed with the subset radius of 10 pixels, the strain window radius of 5 pixels and step size of 4 pixels. Moreover, large strain analysis with an automatic seed propagation was needed. Once the acceptable field characteristics were obtained, they were directly compared with the IR approach with the control point spacing of 60 pixels. Note, that only small changes were obtained with different spacings. A comparison of displacement and strain fields is shown in Fig. 5.

Fig. 5

The distribution of displacements is almost the same which is also confirmed by MAPE of 2-3 % in pixel-to-pixel comparison. The MAPE increases a bit when comparing the strain field. These results are expectable because smoothing is introduced for the local DIC. Nevertheless, the difference between the IR and DIC results is around 4 % only and the strain concentrations and strain distributions are very similar. The structural similarity was very high for all the analysed stages with SSIM values ~ 0.98 indicating a comparable capturing of the deformation field and thus confirming high accuracy of the performed IR analysis.

## 3.3 Skin indentation test with natural texture

Images obtained from the indentation test were firstly modified by cutting off the indenter and the background to improve the contrast and then filtered to highlight the natural structure needed for the analysis. As it was impossible to prepare a suitable speckle pattern for DIC, it was hardly feasible to obtain any results using Ncorr and DIC. Therefore, the performed analyses showed very high correlation coefficients within the ROI, possibly influencing the results. Moreover, some areas were automatically omitted because the correlation coefficient exceeded a cut-off value of the program. Analogous to the uniaxial sample, the displacement field was almost independent on the chosen subset region radius: maximum MAPE was around 5 % (values varied throughout loading stages) when comparing radiuses of 15, 30 and 45 for DIC. However, the impact of different smoothing radius $r_\varepsilon$ on the evaluated strain fields was huge. When a fixed subset radius of 30 was used, the MAPE between $r_\varepsilon = 5$ and 10 was $80 - 170$ % for horizontal strain, $50 - 100$ % for vertical strain and $80 - 130$ % for shear strain. Therefore, it was not possible to obtain a reasonable strain distribution within the skin sample. Consequently, no reliable reference DIC distribution could be used for comparison of IR strain fields and only displacement fields in horizontal and vertical directions were compared (with DIC subset radius of 30, step size 4 pixels).

Spacing between B-spline control points of the IR approach was set to 128 pixels in the horizontal and 64 in the vertical direction. It is important to note, that no straightforward rule could be applied with such a complex sample requiring a proper inspection of the obtained results. A comparison between the resulting full-field displacements is shown in Fig. 6. Except for initial loading stages MAPE was found around 5 % in the horizontal direction and 2 % in the vertical direction (i.e., direction of indentation), showing thus a good agreement with DIC.

Fig. 6

It was important to check also the SSIM as an additional metric for the IR and DIC comparison. Here, the values are around 0.91, which is smaller than for the uniaxial sample but still very high. Moreover, IR managed to keep the same quality even in the areas where DIC was not capable to find any results. Summaries of both MAPE and SSIM metrics are shown in Fig. 7. In the case of SSIM plotted on the whole skin sample, high values can be seen within the region evaluated by the DIC analysis. Nevertheless, some areas show lower values due to presence of a high noise and decrease thus the overall average value. These parts should be treated carefully.

Fig. 7

# 4 Discussion

Characterization of mechanical properties is essential in solid mechanics. Progression of atherosclerosis or formation of an aneurysm increases the tissue heterogeneity which contradicts the assumptions often used in mechanical testing and its subsequent evaluation. Additional information on tissue heterogeneity can help us to improve the tissue characterization. This study shows possible advantages of the proposed IR-based approach for mapping the deformation field in soft tissue mechanical testing. The approach provides an alternative to the well-established DIC, based on the open-source tool Elastix [36], and increases thus its potential exploitation.

A wide application of IR in computer vision, mainly for aligning clinical images of multiple modalities, inspired also the *in vivo* strain analyses between two loading stages [6–8]. As it is impossible to prepare a suitable speckle pattern needed for DIC, a full field strain evaluation depends on a natural pattern only. The results of this study showed that IR is a suitable method in mechanical experiments to obtain rigorous full-field measurements not only within the range of small deformations, but also for large deformations and, the most importantly, with a natural pattern of a biological tissue. High accuracy was obtained for synthetic data but also in real applications found in the field of biomechanics. Fig. 4 and Table 1 show that the specific setup of both IR and DIC can highly influence the resulting fields. However, this is inevitable because some noise and the resulting errors cannot be avoided but only minimized. A recently published study [41] compared different DIC approaches including local and global ones but also a newly proposed augmented Lagrangian DIC method which includes mesh adaptability.

The authors performed also an analysis of a synthetic sample 14 from [40] (used also in this study) and showed a very high error and bias for all the approaches. Our results indicate a very good alignment with the simulated data and extend the results of [34] where the authors used Demons registration for synthetic and experimental data for uniaxial tension tests of steel, though with significant differences from the expected results at strains above 0.05. Comparison of experimental data with unknown deformation fields was done here by using DIC as a reference, although we hoped to show advantages of the proposed IR approach compared to the DIC. As there are considerable differences even among DIC analyses themselves with various parameter settings, some error was expected when validating the IR approach. Nevertheless, DIC is a well-established method and still provides proper information at some parts of the skin sample introduced here and helped us to verify the approach in combination with two metrics of registration quality.

This particular example showed the advantage of our alternative approach for full-field investigations because the missing information when using DIC might be problematic for a subsequent investigation of the tissue constitutive parameters done, for instance, using inverse analysis or virtual fields methods [44,45]. The problem of lost points in DIC could be addressed by a variety of related methods, which are between two extremes proposed here, i.e. the basic DIC technique (no overlap between subsets, rigid displacements in each subset) and the fully global IR technique. In between, it is possible to fit the displacements and their gradients across larger subsets as done in the IR method over large overlapping subsets [46–48]. Nevertheless, also other soft tissues showing a natural pattern may be easily analysed using the proposed IR approach [31–33]. However, an additional check of the registration quality is recommended using another metric such as SSIM used here, otherwise some unreliable deformation can be found. The potential of full-field methods is especially important for diseased soft tissues such as atherosclerotic arteries, the heterogeneity of which disables creation of a regular specimen shape [11,49–51]. In addition to material non-homogeneity, a high variation may result from non-uniform specimen thickness violating the needed assumptions of the test [52]; also these errors can be decreased by using the proposed IR approach.

## 4.1 Limitations

The parameters used in the current work might not be suitable generally for any possible cases as indicated e.g., by the different spacing between B-spline control points. However, the used registration tool Elastix enables an easy configuration of parameters for a wide field of applications. Although an incremental approach was used in this study, updating the undeformed image in every iteration, it might be sometimes useful to have one reference image only. In that case, a different hierarchical strategy (multi-resolution) might be preferred. Moreover, if the error is cumulated through a number of images, it might result in high differences as shown in [34]. Nevertheless, the proposed approach showed a decreasing trend in the differences against the DIC and thus no cumulative effect was suspected.

As we were introducing a proof of concept of the IR methodology, the computational efficiency was out of scope of our study, although it may play an important role. However, it is worth to mention that the performance was better for the Ncorr compared to the Simple-Elastix, although a direct use of Elastix in C++ would probably perform much faster. Another limitation of the proposed approach might be related to discontinuities that are well covered for DIC techniques [53]. Nevertheless, discontinuities might also occur during a patient screening where the movement of organs leads to local discontinuities and thus they should inspire further research in the field of image registration. This problem was recently investigated by using deep learning [54] to obtain better spatial transformation of two image sets without the need for globally smooth and continuous transformation fields. Although incorporation of machine learning is very interesting, the complexity could limit the spread among more users. A more effective way was proposed in [55] who introduced a multi-resolution eXtended Free-Form Deformation inspired by the eXtended finite element method framework used often to analyze discontinuities in DIC-based approaches. The authors used the same non-rigid B-spline image registration adapted through the Elastix as in our study, though they enriched the B-spline essential functions with some extra degrees of freedom near the discontinuity interface. Therefore, they could track the deformation of medical images with discontinuities. The adaptation of this extension can be an exciting motivation for further research when applied to the problems related to the experimental mechanics.

## 5 Conclusion

In this study, we showed that Image Registration can be beneficial for strain measurements in soft biological tissues. We verified the methodology with a challenging benchmark based on simulated data and made proofs of concept on an arterial sample in uniaxial tension and on skin under indentation. Image Registration does not require the creation of an appropriate speckle pattern whereas digital image correlation may fail due to difficulties with the natural pattern. Possible extension of the evaluated region for IR compared to DIC can also be interesting for subsequent identification of constitutive parameters especially for specimens with heterogeneous strain fields where complete full-field measurements are required, even close to the edges. The proposed IR analysis can be applied on a spectrum of biological tissues in the future and even on previous experiments where the recorded images were not useable for standard DIC. Nevertheless, strain fields cannot be transformed directly into stress fields without solving an inverse problem to identify constitutive properties of the non-homogeneous tissue.

# Acknowledgment

This work was supported by Czech Science Foundation project No. 21-21935S.

Fig. 8: A flowchart of IR approach used to map a full-field deformation from experimental measurements. F=0 indicates initial undeformed state, $\mathbf{u}^{full}$ is a matrix containing the calculated displacements, $\mathbf{X}$ is an undeformed state and $\mathbf{x}$ is deformed state of the image T is the transformation and $\mathbf{u}$ is the displacement field.

Fig. 9: Three samples used in the study. A is reused from [40] where artificial heterogeneous displacement field was applied. B is an aorta sample in uniaxial tension and C is a skin sample under indentation test.

Fig. 10: Horizontal displacement and strain of synthetic data Sample 14, L3 from DIC Challenge. An influence of different sizing of control points is shown for (A-B) – 15, (C-D) – 30 and (E-F) – 60. A trade-off between SD and bias to simulated data is required. The absolute bias between the synthetic results and those actually obtained with IR is shown in blue.

Fig. 11: Displacement (left column) and strain (right column) distribution along horizontal direction from the synthetic data of sample 14 with different setups: (A-B) - L1, (C-D) – L3 and (E-F) – L5. The absolute bias between the synthetic results and those actually obtained with IR is shown in blue.

Fig. 12: Comparison of displacement and strain fields obtained using DIC and IR approaches for uniaxial tension in the middle of a loading cycle.

Fig. 13: Comparison between vertical and horizontal displacement fields. DIC (top row) shows a lot of missing data due to a poor correlation coefficient while it was possible to analyse the whole sample with the IR approach (bottom row). To enable straight comparison, the colour bar scale is fixed based on DIC results. Therefore, the maximal vertical displacements in the left bottom figure are beyond the scale of DIC and depicted uniquely in purple.

Fig. 14: Comparison between DIC and IR by means of MAPE (left) and SSIM (right) metrics for the porcine (uniaxial test, in blue) and skin (indentation test, in orange) sample. In the left figure, the solid and dotted lines are related to MAPE for displacements and strains, respectively. Examples of SSIM distribution are shown for both samples indicating noisy response. Some parts, mainly on the edge of the skin sample, show lower values and hence should be treated carefully.